\documentclass{iaus}
\usepackage{graphicx}

\newcommand\ga{\gtrsim}
\newcommand\la{\lesssim}

\def\ee #1 {\times 10^{#1}}          % \ee p       10^p   
\def\ut #1 #2 { \, \textrm{#1}^{#2}} % \ut unit p  unit^p 
\def\un #1 { \, \textrm{#1}}          % \un unit     unit
\def\u #1 { \, \textrm{#1}}          % \un unit     unit
\def\kms {\,\textrm{km\,s}^{-1}}

\title[Keplerian megamaser disks] %% give here short title %%
{The origin of Keplerian megamaser disks}

\author[Mark Wardle \& Farhad Yusef-Zadeh]   %% give here short author list %%
{Mark Wardle$^1$
 \and Farhad Yusef-Zadeh$^2$}

\affiliation{$^1$Department of Physics \& Astronomy and Research Centre for Astronomy, Astrophysics \& Astrophotonics, Macquarie University, Sydney NSW 2109, Australia \\ email: {\tt mark.wardle@mq.edu.au} \\[\affilskip]
$^2$Department of Physics \& Astronomy, Northwestern University,
Evanston IL 60208, USA \\ email: {\tt zadeh@northwestern.edu}}

\pubyear{2012}
\volume{287}  %% insert here IAU Symposium No.
\pagerange{1--2}
\jname{Cosmic Masers -- from OH to H$_0$}
\editors{R. Booth, E. Humphries \& W. Vlemmings, eds.}
\begin{document}

\maketitle

\begin{abstract}
Several examples of thin, Keplerian, sub-parsec megamaser disks have been discovered in the nuclei of active galaxies and used to precisely determine the mass of their host black holes.   We show that there is an empirical linear correlation between the disk radius and black hole mass and that such disks are naturally formed as molecular clouds pass through the galactic nucleus and temporarily engulf the central supermassive black hole.  For initial cloud column densities below about $10^{23.5}$\,cm$^{-2}$ the disk is non-self gravitating, but for higher cloud columns the disk would fragment and produce a compact stellar disk similar to that observed around Sgr A* at the galactic centre.

%% add here a maximum of 10 keywords, to be taken form the file <Keywords.txt>
\end{abstract}

%\firstsection % if your document starts with a section,
              % remove some space above using this command.
%\section{Introduction}

Fourteen Seyfert 2 nuclei are known to host powerful 22\,GHz water masers located in a circumnuclear disk within a parsec of the central massive black hole. In eight edge-on systems, the maser kinematics trace the rotation curve of a thin, Keplerian disk, enabling accurate determination of the black hole mass (Herrnstein et al.\ 2005; Kuo et al.~2010).   22\,GHz water masers occur where the disk surface density exceeds $1\un g \ut cm -2 $ and warping exposes the surface to X-ray irradiation from the centre (Neufeld et al.~1994; Maloney 2002).  
We propose instead that Keplerian megamaser disks are created through the partial capture of molecular clouds (Wardle \& Yusef-Zadeh 2012), a model we have previously used to explain the compact disk of stars orbiting within 0.1\,pc of Sgr A* in our own Galactic centre (Wardle \& Yusef-Zadeh 2008). In this scenario a dense molecular cloud sweeps through the central few parsecs, temporarily engulfing the massive black hole.  Streams of material passing on opposite sides of the hole are gravitationally deflected towards each other and collide, dissipating kinetic energy and partly cancelling angular momentum.  The shocked gas cools efficiently, circularises, and forms a compact, thin molecular disk.

The disk size is determined by the black hole mass $M$, the initial speed $v$ of the incoming cloud and the effectiveness of angular momentum cancellation.  We make 2 simple assumptions:
cloud material with initial impact parameter less than $r_\mathrm{acc} = 2GM/v^2$ is captured; and the angular momentum of this material is reduced to a fraction $\lambda$ of its initial magnitude.
The radius $R$ of the disk is estimated by equating the specific orbital angular momentum at the disk edge, $\sqrt{GMR}$, to that of the material that is just barely captured, i.e.\ $\lambda r_\mathrm{acc} v$\,; this yields $ R = 4G\lambda^2 M / v^2 $.
For $v = 225\kms$ and $\lambda=0.3$ the predicted disk radius is 0.3\,pc for a $10^7$\,M$_\odot$ black hole and matches the empirical linear relationship of maser disk size with black hole mass (see left panel of Fig. 1).
\begin{figure}
\begin{center}
      \includegraphics[scale=0.49]{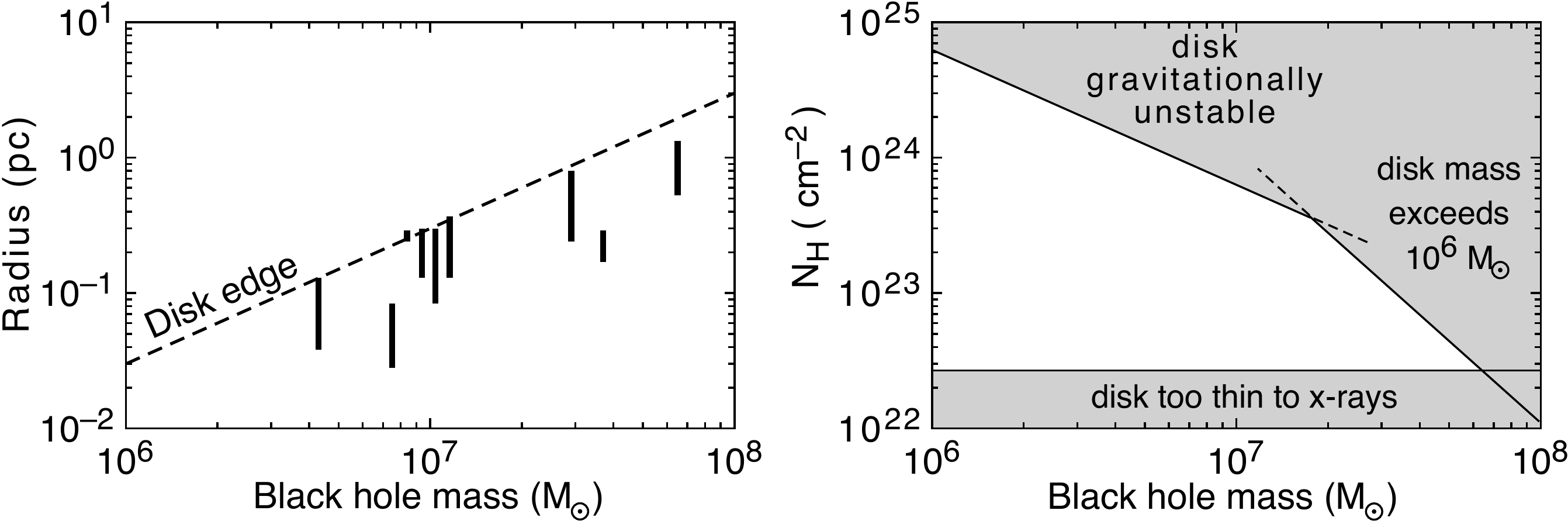}
\caption{\emph{(Left)} Vertical bars: observed radial extent of Keplerian rotation curves traced by megamasers in Seyfert 2 nuclei.  Dashed curve: predicted disk edge in the cloud capture scenario.    From left to right the systems are  Sgr A* (compact stellar disk, Lu et al.~2009); NGC 2273, NGC 4388, NGC 6323, UGC 3789, NGC 2960, and NGC 6264 (Kuo et al.~2011), NGC 4258 (Herrnstein et al.~2005), and NGC 1194 (Kuo et al.~2011).
\emph{(Right)} Constraints on cloud column density as a function of black hole mass.  The column density must lie in the unshaded region to produce a thin, Keplerian disk hosting 22\,GHz megamasers.}
\end{center}
\end{figure}

The disk surface density $\Sigma_D$ can be estimated from the captured mass $\sim \pi r_\mathrm{acc}^2 \Sigma_\mathrm{cloud} $, where $\Sigma_\mathrm{cloud}$ is the cloud surface density.  $\Sigma_D$ must meet two conditions to host  22\,GHz masers:
$\Sigma_D \ga 1$\,g\,cm$^{-2}$ so that x-ray irradiation produces a warm, dense, H$_2$O-rich layer; and $\Sigma_D\la c_s\Omega/\pi G$, where $c_s \approx 1$\,km\,s$^{-1}$ is the sound speed in the layer, so that the disk is not self-gravitating. In addition the inferred disk mass should neither exceed the notional cloud mass nor be sufficient to distort the rotation curve.
The corresponding constraints on the column density of the incoming cloud are shown in Fig. 2.

This partial capture scenario naturally produces megamaser disks on sub-parsec scales, with the observed linear correlation between disc radius and black hole mass. The outer edge of the maser discs is due to physical truncation rather than a breakdown in masing conditions within an extended disk.   This picture relies on dense clouds occasionally sweeping through the inner few parsecs of galactic nuclei. The circumnuclear molecular ring at 1.7\,pc from Sgr A* (e.g.\ Christopher et al. 2005), and  the 10\, pc scale circumnuclear rings in numerous Seyfert galaxies suggest that there is an ample supply of material from larger radii, perhaps controlled by angular momentum transfer between massive clouds during gravitational encounters (Namekata \& Habe 2011).  The formation of gravitationally \emph{unstable} disks may be a more common event because molecular clouds in the inner regions of galaxies tend to have column densities $\ga 10^{24}\ut cm -2 $.  These transient disks may also host megamasers (Milosavljevi\'c \& Loeb 2004), so that only a small fraction of megamaser AGN may have the Keplerian disks necessary for accurate black hole mass determinations.

\newcommand\refitem{\bibitem[]{}}
\newcommand{\ApJ}{\textit{ApJ }}
\newcommand{\ApJL}{\textit{ApJ} (Letters) }
\newcommand{\PASA}{\textit{PASA }}

\end{document}